\documentstyle[floats,aps,prl,epsf]{revtex}
\begin{document}
\draft

\newcommand{\figellpara}{
\begin{figure}[h]
	\setlength{\unitlength}{1mm}
	\begin{picture}(90,90)(0,0)
	\put(66.5,80){\makebox{$\scriptstyle 1$}}
	\put(49,73){\makebox{$\scriptstyle 2$}}
	\put(64,13){\makebox{$\scriptstyle L-1$}}
	\put(43,51){\makebox{$\scriptstyle 1$}}
	\put(43,41){\makebox{$\scriptstyle L-1$}}
	\put(-10,0){
        	\makebox{
			\setlength\epsfxsize{9cm}
                	\epsfbox{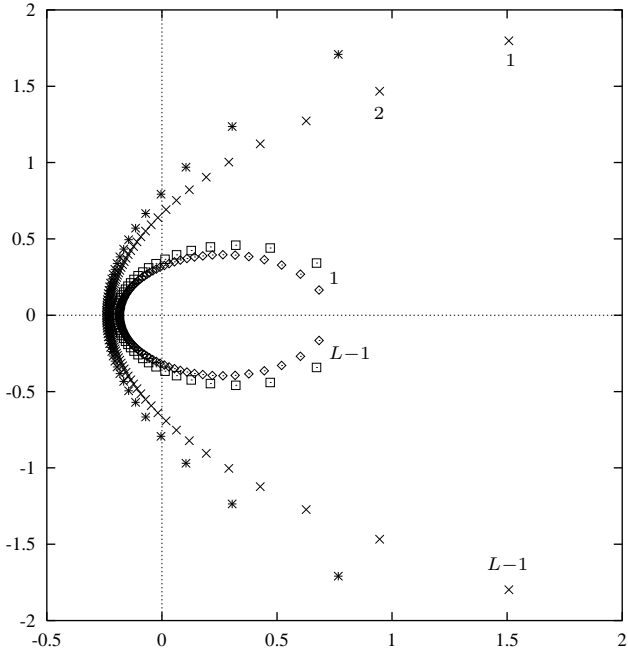}}}
	\end{picture}
\caption{
The roots of $Z$ in the complex $x$-plane. Shown are the $L-1$ roots for
$q=2.0$: $L=50$ (squares) and $L=100$ (diamonds); 
$q=2.5$: $L=50$ (stars) and $L=100$ (crosses). 
        }
\label{figellpara}
\end{figure}
}
%
%
\newcommand{\figmpamf}{
\begin{figure}[h]
	\setlength{\unitlength}{1mm}
	\begin{picture}(90,80)(0,0)
	\put(-1,73){\makebox{$\tilde{\rho}$}}
	\put(77,+2){\makebox{$q$}}
	\put(-10,0){
        	\makebox{
			\setlength\epsfxsize{9cm}
                	\epsfbox{fig2.epsf}}}
	\end{picture}
\caption{
The phase diagram in the $\rho$-$q$-plane. The crosses mark the
phase transition obtained by the described method, the
thin line results from mean-field considerations. 
        }
\label{figmpamf}
\end{figure}
}
%
%
\newcommand{\figden}{
\begin{figure}[h]
	\setlength{\unitlength}{1mm}
	\begin{picture}(90,80)(-3,0)
	\put(-6,73){\makebox{$g$}}
	\put(62,+3){\makebox{$\tau-l/2$}}
	\put(-15,0){
        	\makebox{
			\setlength\epsfxsize{9cm}
                	\epsfbox{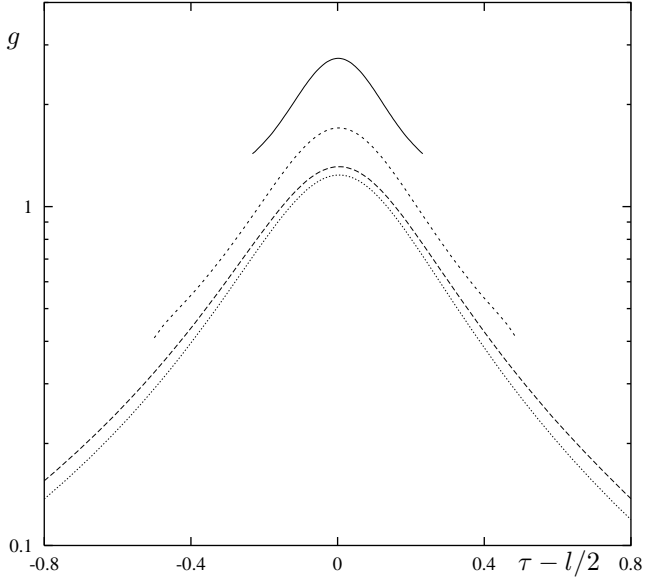}}}
	\end{picture}
\caption{
The density of roots as a function of the arc length for 
$L=100$ and $q=1.2,\,1.5,\,2.0,\,2.5$ (from top to bottom).
The curve of roots in the $x$-plane is ellipsoidal in the first three cases 
and hyperbolic in the last one.
The curves are shifted by half of their arclength $l$.
        }
\label{figden}
\end{figure}
}
%
%
\newcommand{\figZ}{
\begin{figure}[h]
	\setlength{\unitlength}{1mm}
	\begin{picture}(90,80)(-3,0)
	\put(50,64){\makebox{$P$}}
	\put(55,39){\makebox{$\rho$}}
	\put(70,+2){\makebox{$x$}}
	\put(-15,0){
        	\makebox{
			\setlength\epsfxsize{9cm}
                	\epsfbox{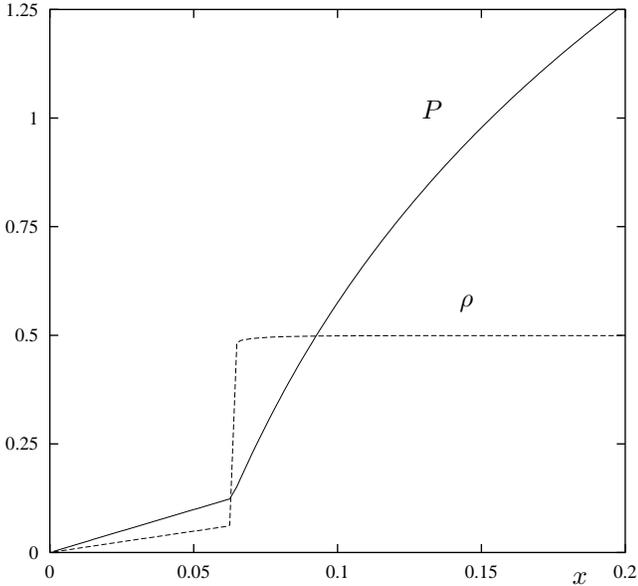}}}
	\end{picture}
\caption{
The ``pressure'' $P=(1/L)\log  Z$ and the density $\rho$ as a function of the fugacity
for $q=1.2$ and $L=400$.
        }
\label{figZ}
\end{figure}
}
%
%
\newcommand{\figfilm}{
\begin{figure}
	\setlength{\unitlength}{1mm}
	\begin{picture}(90,70)(0,0)
	\put(8,2){
        	\makebox{
			\setlength\epsfxsize{6cm}
                	\epsfbox{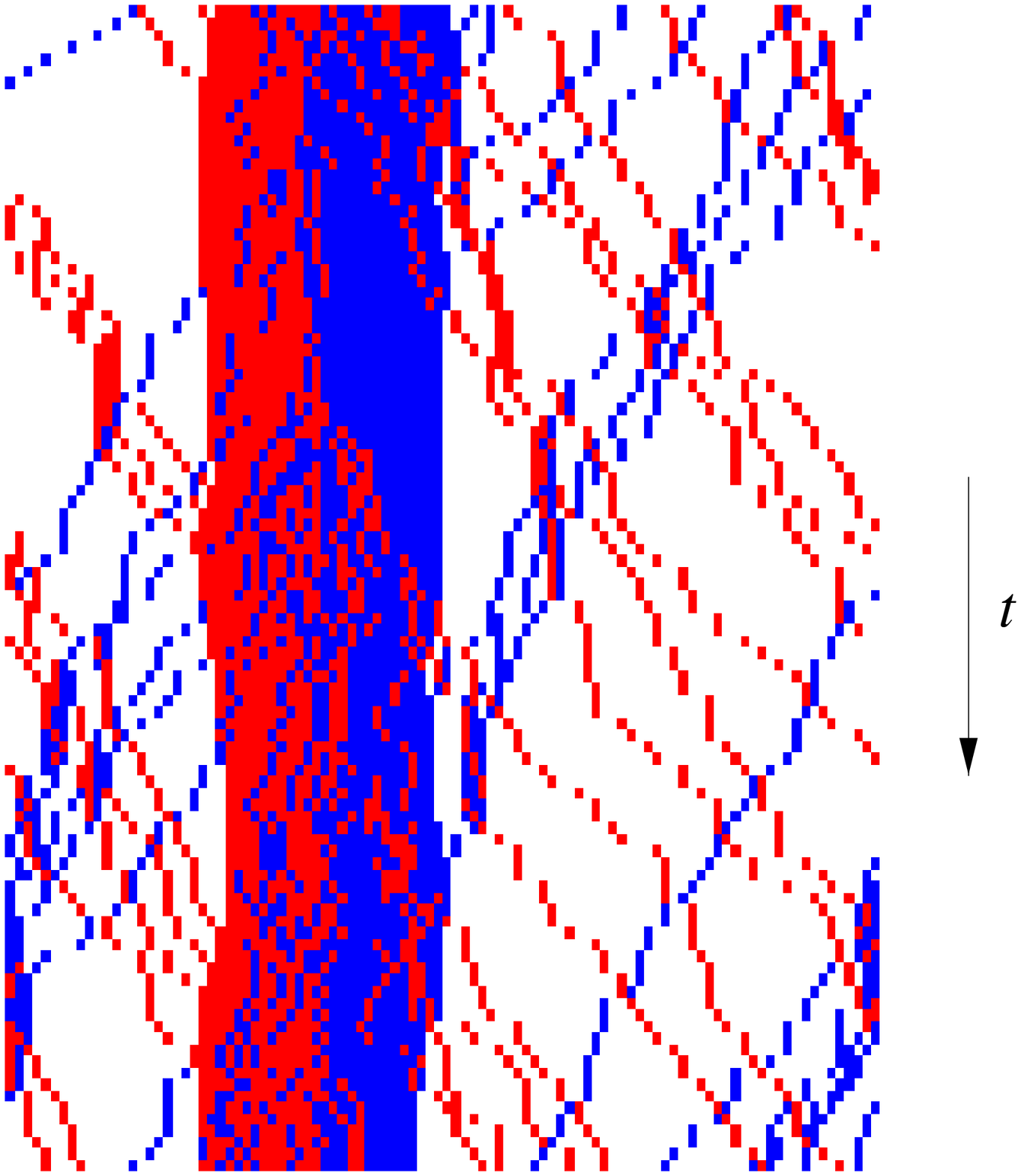}}}
	\end{picture}
\caption{
A film of a Monte-Carlo simulation of the model for 
$q=1.2$, $\rho_1=\rho_2=0.2$ and $L=100$. 
Particles of type 1 are grey and those of type 2 black.
        }
\label{figfilm}
\end{figure}
}
%
%
\twocolumn[\hsize\textwidth\columnwidth\hsize\csname@twocolumnfalse%
\endcsname

\title{Yang--Lee Theory for a Nonequilibrium Phase Transition}
\author{Peter F. Arndt}
\address{
Institut f\"ur Theoretische Physik,
Universit\"at zu K\"oln,
Z\"ulpicher Strasse 77,
50937 K\"oln, Germany}

\date{January 13, 2000}
\maketitle

\begin{abstract}
To analyze phase transitions in a nonequilibrium system we study its grand
canonical partition function as a function of complex fugacity.  Real and
positive roots of the partition function mark phase transitions.  This
behavior, first found by Yang and Lee under general conditions for equilibrium
systems, can also be applied to nonequilibrium phase transitions.  We consider
a one-dimensional diffusion model with periodic boundary conditions.
Depending on the diffusion rates, we find real and positive roots
and can distinguish two regions of analyticity, which can identified with  two
different phases.  In a region of the parameter space both of these phases
coexist.  The condensation point can be computed with high accuracy.
\end{abstract}

\pacs{PACS: 
	05.20.-y, 
	02.50.Ey, 
	05.70.Fh, 
	05.70.Ln 
	}

]

The investigation of nonequilibrium systems is a growing field in statistical
mechanics and currently attracts much attention.  In this context simple models
such as driven diffusive systems play a paradigmatic role similar to the Ising
model in equilibrium statistical mechanics.  These systems establish a simple
framework in which many phenomena can extensively be studied.  Moreover, driven
diffusive systems can easily be mapped to other nonequilibrium models, e.g.
of polymer dynamics, interface growth and traffic flow
\cite{PrBook,ScZi,HaZh,Na}.

A hallmark of many nonequilibrium systems is the absence of detailed balance
and the support of stationary states with non-vanishing currents.  Hence these
systems build a larger class than respective equilibrium systems and phase
transitions may appear under less restrict conditions.  For example, it is
known that spontaneous symmetry breaking and a first-order phase transition may
occur in one-dimensional nonequilibrium systems with short-range interactions
\cite{SSB}.

In thermal equilibrium the probability measures can in principle be expressed
through an appropriate ensemble.  For driven systems an equally powerful
concept is missing.  In Ref. \onlinecite{DeJaLeSp} a grand canonical partition
function for nonequilibrium systems has been introduced for the first time.
For the definition of this function one uses the matrix-product representation
(see below) of the stationary state. Such a representation is not known for
every system. But for several models it is known and, for instance, its
existence is guaranteed for open reaction-diffusion models \cite{KrSa}.
In this Letter we utilize the partition function and show that more concepts
from equilibrium physics may be applied to nonequilibrium systems.  We develop
a Yang--Lee theory \cite{YaLe} giving us a very powerful method to analyze
phase transitions.

We start from the grand canonical partition function $Z(x)$ and study its
behavior as a function of complex fugacity $x$.  Although only real values of
the fugacity are of physical interest, the analytical behavior of thermodynamic
functions can be revealed completely only by allowing the fugacity to be
complex.  This was first found by Yang and Lee for equilibrium systems
\cite{YaLe}:  For finite systems the roots of the grand canonical partition
function $Z(x)$ are in general complex or negative if real.  But in the
thermodynamic limit roots  may approach the positive real axis.  This marks a
phase transition; in equilibrium systems the pressure $p=k_{\rm B} T
\lim_{V\rightarrow\infty} ((1/V) \log Z)$ is non-analytical, the density
$\rho=(\partial/\partial\log x)(p/k_{\rm B}T)$ is discontinuous,  and one can
distinguish different phases.

We show here that the above behavior of the roots of the partition function can
also be found for nonequilibrium systems.  By similar reasoning, one can
clearly define phases and determine their properties and furthermore transfer
other concepts known from equilibrium statistical mechanics to nonequilibrium
systems.  A new method to compute the phase transition point with great
accuracy is established, as well.

\medskip

To present our results let us consider an one-dimensional stochastic diffusion
model with $L$ sites and periodic boundary conditions.  Each site $k$ may be
occupied by one particle of type 1 or 2, or may be vacant (denoted by 0).
Starting from random initial conditions with fixed particle densities, $\rho_1$
and $\rho_2$, the particles can rearrange themselves.  The only allowed
processes are local interchanges conserving the number of particles:
$(\alpha)_k(\beta)_{k+1}\rightarrow (\beta)_k(\alpha)_{k+1}$ with $\alpha,\beta
\in \{0,1,2\}$.  In an infinitesimal time interval $d\tau$ these diffusion
processes occur with probability $g_{\alpha,\beta}\, d\tau$.  Here we consider
the rates:
\begin{equation}
g_{1,2}=q,\quad
g_{2,1}=1,\quad
g_{1,0}=g_{0,2}=1
\end{equation}
\def\qcr{q_0}%
All other $g_{\alpha,\beta}$ are zero.  The stationary state of this model has
been studied recently by Monte-Carlo simulations and mean-field approximations
\cite{ArHeRiModel}.  For a snapshot of a Monte-Carlo simulation see Fig.
\ref{figfilm}.  For the parameters chosen one observes one macroscopic droplet
of particles which is surrounded by vacancies with a finite density of
particles.  This is a first sign of a first order phase transition.  Increasing
$q$ the droplet shrinks and disappears for $q$ larger than some $\qcr(\rho)$;
decreasing $q$ it grows until, for $q=1$, all particles are bound in the
droplet.  For $q>1$ the position of droplet fluctuates and is fixed for $q<1$,
i.e. translational invariance is spontaneously broken.  Similar one- and
higher-dimensional models have been studied in \cite{EvKaKoMu,ScZi}.

\figfilm

The probability, ${\cal P}(\bbox{\beta},t)$, for configurations
$\bbox{\beta}=(\beta_1,\beta_2,\dots,\beta_L)$ at time $t$ follows a Master
Equation \cite{PrBook}.  The stationary probability distribution 
$\cal P_{\rm st}(\bbox{\beta})$ can be expressed as the trace over
algebra elements $D_\beta$ \cite{ArHeRiAlgebra}:
\def\tr{{\rm tr}}
\begin{equation}
{\cal P}_{\rm st}(\bbox{\beta})=
\frac1Z \tr (D_{\beta_1} D_{\beta_2} \cdots D_{\beta_L})
\label{eqstat}
\end{equation}
The normalization factor, $Z$, is defined as
\begin{equation}
Z=\sum_{\beta_1,\beta_2,\dots,\beta_L}  
\tr (D_{\beta_1} D_{\beta_2} \cdots D_{\beta_L})
=\tr (C^L)
\end{equation}
with $C=D_0+D_1+D_2$. 
In the following we recognize that $Z$ plays the role
of a grand canonical partition function.
For Eq. (\ref{eqstat}) to hold,
the $D_\beta$ have to fulfill the quadratic algebra 
\def\G{{\cal G}}
\def\Gn{\G_0}
\def\Gi{\G_1}
\def\Gii{\G_2}
\begin{mathletters}
\label{Dalg}
\begin{equation}
q D_1 D_2 - D_2 D_1=x_2 D_1 + x_1 D_2
\end{equation}
\begin{equation}
D_1 D_0 = x_1 D_0
\end{equation}
\begin{equation}
D_0 D_2 = x_2 D_0
\end{equation}
\end{mathletters}
The two free parameters $x_1$ and $x_2$ reflect the freedom 
to choose the two densities of particles.
A representation of this algebra is known \cite{ArHeRiModel}
\begin{equation}
D_0 = \Gn                   \; , \quad
D_1 = x_1 \Gi             \; , \quad
D_2 = x_2 \Gii
\end{equation}
with the matrices $\G$ given by

\[
(\Gn)_{ij}=\delta_{1i}\delta_{1j}
\]
\[
\Gi\!=\! \left(\begin{array}{lllll}
                a_1&t_1&0&0&\!\!\cdots        \\
                0&a_2&t_2&0&              \\
        \vspace*{-2mm}
                0&0&a_3&t_3&              \\
                0&0&0&a_4&\!\!\ddots           \\
                \vdots&&&&\!\!\ddots
            \end{array}\right)
,\,
\Gii\!=\! \left(\begin{array}{lllll}
                a_1&0&0&0&\!\!\cdots        \\
                s_1&a_2&0&0&              \\
                0&s_2&a_3&0&              \\
                0&0&s_3&a_4&              \\
                \vdots&&&\ddots&\!\!\ddots
            \end{array}\right)
\]
where we have introduced the notations
\[
a_k     =r\left( 2 \{k-1\}_r - \{k-2\}_r \right)
\]
\[
s_k\, t_k=\{k\}_r\, 
\left( 3r-1+(2 r-1)^2 \,\{k-1\}_r \right)
\]
\[
\{k\}_r=\frac{r^k-1}{r-1}
,\quad
r=\frac1q
\]

In the following
we restrict ourselves to equal densities of particles, 
$\rho_1=\rho_2=\rho$, and have $x_1=x_2=x$. Then
\begin{equation}
Z=Z(x)=\sum_{\beta_1,\beta_2,\dots,\beta_L} 
x^{N(\bbox{\beta})} \tr (\G_{\beta_1} \G_{\beta_2} \cdots \G_{\beta_L})
\end{equation}  
takes the form of a grand canonical partition function where $N(\bbox{\beta})$
counts the number of particles in a configuration $\bbox{\beta}$ and 
therefore $x$ can be
interpreted as a fugacity.  The density of particles 1 is given by
\begin{equation}
\rho(x)
=\frac1{Z} \tr (D_1 C^{L-1})
=\frac12 \frac{\partial}{\partial \log x} P(x)
\label{eqrho}
\end{equation}  
For the last expression in Eq. (\ref{eqrho}) we introduced, analogously to
equilibrium physics, the ``pressure''
\begin{equation}
P(x)=\frac1L \log Z(x)
\end{equation}  
Note, that in this context $P$ is not the physical pressure of the particles.
The singularities and the analytical behavior of $\rho(x)$ and $P(x)$ are 
determined by the roots of $Z(x)$.  To perform a numerical calculation of $\tr
(C^L)$ we change the ensemble slightly and require that at least one vacancy is
present.  In this case we have to compute $\tr (\Gn C^{L-1})=(C^{L-1})_{11}$
which is much simpler and one has to handle $(L/2)\times(L/2)$ matrices only.
Accordingly, 
\begin{equation}
Z(x) \propto \prod_{j=1}^{L-1} (x-x_j)
\end{equation}  
is a polynomial of degree $L-1$ in $x$ with roots $x_j$. 

\figellpara

\def\xcr{\tilde{x}}%
\def\rhocr{\tilde{\rho}}%
To study the analytical behavior of $P$ and $\rho$ we first determine the roots
of $Z(x)$ in the complex plane.  Since the partition function is a real
polynomial with positive coefficients the roots come in complex conjugated
pairs and real roots are negative.  The roots of $Z$ for two different choices
$q>1$ are shown in Fig.\ \ref{figellpara}.  For small $q$ the roots are lying
on an elliptic curve.  For $q$ larger than a critical value, $q_{\rm crit}
\approx 2.15$, the curve is a hyperbola.

Let us treat the elliptic before the hyperbolic case.  At the end we give some
remarks on the case $q<1$.

{\em The elliptic case} ($1<q<q_{\rm crit}$). ---
In this case, the imaginary part of the root closest to the positive real axis
(e.g. root No. 1 for $q=2.0$ in Fig.\ \ref{figellpara}) vanishes for large $L$
with $L^{-1}$  while the real part of this root stays
finite and reaches a positive value, $\xcr(q)$, in the limit $L\rightarrow\infty$.

\figden

To investigate further, let us denote the curve from the first to the last root
(see Fig.\ \ref{figellpara}) by $c(\tau)$ with $\tau \in [0,l]$ where $l$ is the
arclength of $c$.  The roots are not equally spaced and their density varies
with the position on the ellipse.  Furthermore, with increasing system size the
density of roots along the curve grows linearly in leading order and we can
denote the number of roots in a line element $d \tau$ by $g(\tau)\, L\, d\tau$.
The function $g(\tau)$ has a non-vanishing large $L$ limit and is shown in
Fig.\ \ref{figden}.  Since in this limit the curve $c(\tau)$ is as well defined
we find the pressure (up to an additive constant)
\begin{equation}
P(x)=\int \!d \tau\, g(\tau)\, \log (x-c(\tau))
\end{equation}
It is a thermodynamic quantity in the sense that its limit
$L\rightarrow\infty$ exists. 
If $q<q_{\rm crit}$ the curve $c$ closes to give an ellipse with a positive
intercept of the real axis: $c(0)=\xcr>0$ and $g(0)>0$ in the limit
$L\rightarrow\infty$.  One has to distinguish two regions of analyticity, the
inner and the outer of the ellipse in the complex plane.  In both regions all
physical quantities are continuous and differentiable.  For a plot of $P(x)$
and $\rho(x)$ see Fig.\ \ref{figZ}.

\figZ

In the outer region of the ellipse ($x>\xcr$) the asymptotic behavior of $P$ is
$P(x)=\log x$ and we have $\rho(x)=1/2$ in this phase.

For $x$ in the inner of the ellipse we find a $\rhocr$ with 
$\rho(x)<\rhocr<1/2$.  Crossing the
ellipse at $\xcr$ the pressure $P(x)$ is not differentiable and $\rho(x)$ has a
finite discontinuity; the transition is of first order. The jump in the density
is related to the density of roots $g(0)$ at the real axis \cite{YaLe}
\begin{equation} 
1/2-\rhocr=\pi \xcr g(0) 
\label{eqrhog} 
\end{equation}
The value of $\xcr$ can easily be obtained by extrapolating the real part of
the first root for $L\rightarrow\infty$. Using Eq. (\ref{eqrho}) and extrapolating 
one gets $\rhocr$.

If the density $\rho$ is fixed in the interval
$\rhocr<\rho<1/2$ one finds the two phases coexistent in the stationary state;
the dense one with $\rho=1/2$ (i.e. without vacancies) and the other with
$\rho=\rhocr$ (see again Fig.\ \ref{figfilm}).  The dense phase acts like a
bottle-neck and accordingly the current is $(q-1)\rho(1-\rho)=(q-1)/4$.

Increasing $q$ to $q_{\rm crit}$, the density of particles above that condensation
starts, $\rhocr$, reaches $1/2$ and the density of roots on the positive real
axis $g(0)$ vanishes according to Eq. (\ref{eqrhog}).  The phases in the inner
and outer region become similar and one finds a second order phase transition
at $q=q_{\rm crit}$.

Next we want to fix the density of particles and determine the value
$\qcr(\rho)$ of $q$ below which two phases coexist.  Therefore, the values of
$\rhocr$ are plotted versus $q$ in Fig.\ \ref{figmpamf}.  In the same plot we
give the mean-field approximation for the phase transition $\qcr^{\rm
MF}(\rho)=(1+6\rho)/(1+2\rho)$ as it has been calculated in \cite{ArHeRiModel}.
In the latter paper the case $\rho=0.2$ was studied by Monte-Carlo simulations
as well and the phase transition was found at $\qcr^{\rm MC}=1.62 \pm 0.05$.
With the above method we find the more precise value $\qcr(0.2)=1.617\pm
0.001$.

\figmpamf

{\em The hyperbolic case} ($q>q_{\rm crit}$). --- 
Also shown in  Fig.\ \ref{figellpara} are the roots for $q=2.5$.  
The curve $c(\tau)$ is a hyperbola.  The distribution of roots
$g(\tau)$ (see Fig.\ \ref{figden}) is similar to the elliptic case.  The real
and imaginary part of the first and last roots, and hence the arclength, $l$,
diverges with $L$.  We find only the phase with $0<\rho(x)<1/2$ continuous and
monotonically increasing with $x$. 

{\em The case $q<1$.} ---
In this case the roots lie nearly equally spaced on a circle.
But the diameter of the circle shrinks exponentially with $L$
and in the large $L$ limit all roots are located at $x=0$.  In the limit
$L\rightarrow\infty$ we find $\rho(x)=1/2$ for $x>0$.  Details on the
$L$-dependence and geometric properties of $c(\tau)$ for all above cases will
be given elsewhere \cite{tocome}.
In physically interesting cases the density of particles is fixed in the
interval $0<\rho<1/2$ and again one finds again coexistent phases. 

\medskip

In summary, we have demonstrated that by allowing for complex values of the
fugacity in the grand canonical partition function one can analyze the phases
and phase transitions in nonequilibrium models.  In the model presented, a
first order phase transition has been investigated.  In particular, the
transition point can be computed with great accuracy.

Due to the general nature of the method it can also be applied numerically or
analytically to other models \cite{tocome} where the density of particles
controls a phase transition, e.g. jamming transitions in traffic related
models.  Furthermore, the method is not constrained to investigate roots in the
fugacity plane. One can study complex roots of some other parameter (e.g. the
input rate in the totally asymmetric exclusion process \cite{DeEvHaPa}) as well
\cite{tocome}.

I want to thank M.R. Evans, C.~Godr\`{e}che, T. Heinzel, D. Mukamel and V.
Rittenberg for discussions.

\end{document}